\documentclass[%
reprint,
twocolumn,
superscriptaddress,
showpacs,preprintnumbers,
nofootinbib,
aps,
prl,
]{revtex4-1}

\usepackage{subfigure}

\usepackage{amsmath}
\usepackage{amssymb}
\usepackage{mathtools}

\usepackage{hyperref}

\usepackage{graphicx}
\usepackage{color}

\usepackage{ulem}

\usepackage{comment}

\graphicspath{
  {./}
  {./figures/}
}

\newcommand{\beq}{\begin{equation}}
\newcommand{\eeq}{\end{equation}}
\newcommand{\bea}{\begin{eqnarray}}
\newcommand{\eea}{\end{eqnarray}}
\newcommand{\ba}{\begin{array}}
\newcommand{\ea}{\end{array}}
\newcommand{\bit}{\begin{itemize}}
\newcommand{\eit}{\end{itemize}}

\newcommand{\eq}[1]{Eq.~(\ref{#1})}

\definecolor{purple}{rgb}{0.5,0,0.5}

\begin{document}
\newcommand{\TQ}{\affiliation{
MOE Key Laboratory of TianQin Mission, TianQin Research Center for Gravitational Physics \& School of Physics and Astronomy, Frontiers Science Center for TianQin, CNSA Research Center for Gravitational Waves, Sun Yat-sen University (Zhuhai Campus), Zhuhai 519082, China.
}}

\title{Domain wall networks from first-order phase transitions and gravitational waves}

\author{Dongdong~Wei}
\email{weidd5@mail2.sysu.edu.cn}
\author{Yun~Jiang}
\email{jiangyun5@sysu.edu.cn}
\TQ

\date{\today}

\begin{abstract}
In the first-order phase transitions (PTs) colliding bubble is an important gravitational wave (GW) source. Following bubble collision, domain walls can be formed when degenerate vacua occur as a result of the breaking of a discrete symmetry relevant to new physics at electroweak or higher scales. Using lattice simulations, we study the dynamical evolution of domain walls and find that the networks of the domain wall are formed around the completion of PTs and the lifetime of the wall networks largely depends on whether or not the degeneracy of true vacua is broken. Our numerical results indicate that domain wall networks continue to produce GWs in the aftermath of PTs, leading to dramatically changing the spectral shape and enhancing the magnitude by about one order. The resulting GW power spectra are peaked at $kR_* \simeq \pi$, above the peak wavenumber it has a decaying power law close to $k^{-1.2}$ followed by a slowly decreasing plateau with the UV cutoff at $kR_* \sim \mathcal{O}(10^2)$.
\end{abstract}

\maketitle
 
\emph{Introduction.} 
The domain wall is one of the important topological defects. It is a product of the breaking of a discrete symmetry, which can be accomplished via cosmological phase transitions (PTs). The traditional mechanism where domain walls are produced from second-order or smooth PTs was originally proposed by T. D. Lee~\cite{Lee:1974jb} and has been studied in the context of $\mathbb{Z}_2$ symmetry~\cite{Bernal:2015xba,Babichev:2021uvl} that is particularly interesting for building a dark matter candidate. Very recently Ref~\cite{Zhou:2020ojf} discusses the formation of the collapsing domain walls in the models with $\mathbb{Z}_3$ symmetry and beyond~\cite{Zhou:2020ojf,Wu:2022stu,Wu:2022tpe}.
Nonetheless, in the early universe the first-order PT is a more attractive scenario as it provides an interpretation of the observed baryon asymmetry of the universe~\cite{Sato:1981ds}. 
To generate a first-order PT, the scalar sector of the potential needs to be constructed such that a barrier separating two degenerate minima appears at the critical temperature. 
Such a construction can be achieved by introducing new particles interacting with the standard model Higgs boson $H$ or introducing a high-dimensional operator $(H^\dagger H)^3$~\cite{Bodeker:2004ws,Grojean:2004xa}. 

Given a discrete symmetry, the first-order PT leads to multiple stable minima (called true vacua) that break the discrete symmetry but have the same potential energy,
while the old vacuum preserving the symmetry before the PT becomes quantum mechanically metastable, which would tunnel to the true vacua through the nucleation of handful of bubbles of true vacuum within the surrounding sea of old vacuum.
Since none of these true vacua is thermodynamically favorable, the scalar field stays at one of the true vacua in the interior of the bubble. 
Following nucleation, these bubbles promptly expand until their walls move at a speed close to the speed of light. 
More realistically, the bubbles interact with the cosmic fluid, resulting in friction on the bubble wall. In this work, we consider the simplest case of bubbles in a vacuum environment. 
As the bubbles continue to expand by many orders of magnitude, more and more regions of the universe convert to true vacuum states. When these different regions are close together, a smooth transition will be made between different true vacua, and the transition region is called the domain wall.

According to the Kibble mechanism~\cite{Kibble:1976sj}, outside a Hubble region the PT proceeds independently and the field configuration may be uncorrelated. Hence, the universe may end up with domains of different vacua divided by a large network of domain walls. 
Unlike cosmic strings, the existence of domain walls is cosmologically catastrophic due to the fact that once produced, their energy density would soon dominate the energy density of the universe, and also would lead to excessive fluctuations in the temperature of the cosmological microwave background (CMB)~\cite{Lazanu:2015fua,Kolb:1990vq}. This is the so-called domain wall problem. To escape from this problem, there must be a mechanism that prevents the production or rapidly destroy them if produced. The solutions to the latter one involve explicitly breaking the discrete symmetries~\cite{Zel'dovich:411756,Gelmini:1988sf} and introducing biased conditions in the initial formation~\cite{Coulson:1995nv}. 

In the vacuum environment only colliding bubbles are the gravitational wave (GW) source during the first-order PTs. As vacuum bubbles collide, a large amount of the energy that was deposited in the bubble wall produces GWs~\cite{Lewicki:2020azd,Lewicki:2019gmv,Ellis:2019oqb,Ellis:2020nnr}. 
For the model where domain walls are seeded by the vacuum bubbles, the generation mechanism of GWs may be more complicated. Although the power spectra of the GWs produced from bubble collisions and the decay of domain walls have been individually studied~\cite{Cutting:2018tjt,Child:2012qg,Kosowsky:1992vn,Hiramatsu:2010yz,Saikawa:2017hiv,Gleiser:1998na}, it is, however, short of relevant studies to investigate the properties of the formed domain walls as well as to make the prediction for the produced power spectrum of the GWs. Also, it is unconfirmed that the total power spectrum is just an incoherent superposition of the one from two sources, particularly for the case where their length scales are close. 

For both purposes, one needs to acquire the dynamics of the scalar field from the equation of motion, which is difficult to solve due to its nonlinear nature. In this work we follow Ref~\cite{Cutting:2018tjt} to implement lattice simulations based on the \textsc{LATfield2} library~\cite{Daverio:2015ryl}. Our numerical results indicate that the GW production does not cease after the PT but lasts until the wall network decay away, subsequently propagating as a stochastic background to the present day. The obtained power spectrum has a very different spectral shape with the amplitude enhanced by about one order at the peak frequency. 

\begin{figure}[t]
\includegraphics[width=0.45\textwidth]{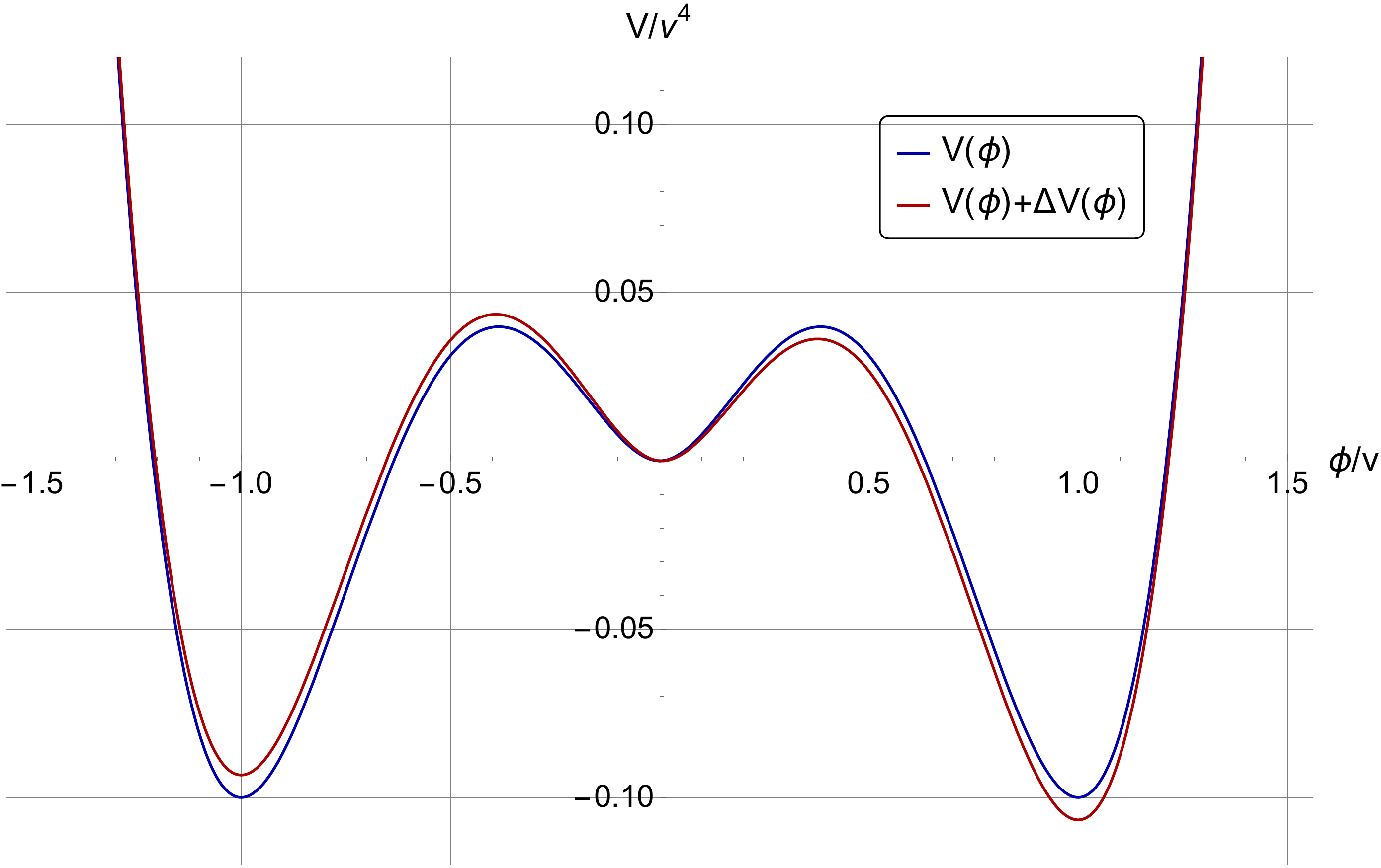} 
\caption{Two models: the blue line draws the potential of \eq{eq:potential} for $a=1$ and the red line additionally include the bias term of \eq{eq:biaspot} with $\epsilon=0.01$.}
\label{fig:potential}
\end{figure}

\emph{A model of domain wall formation}. 
To demonstrate the formation of the domain walls we consider the spontaneous breaking of $\mathbb{Z}_2$ symmetry.
As an example of the scalar model leading to the first-order PTs, let us use a polynomial potential in terms of $|\phi|$ (see Fig.~\ref{fig:potential}),
 \begin{equation}
V(\phi)= a v^{2}|\phi|^{2}-(2 a+0.4) v|\phi|^{3}+(a+0.3)|\phi|^{4}
\label{eq:potential}.
\end{equation}
This model has a $\mathbb{Z}_2$ symmetry $\phi \to - \phi$ with two global minima (true vacua) at $\phi = \pm v$, while $\phi=0$ is a local minimum (metastable vacuum).

Suppose the scalar field $\phi$ is initially trapped in the vicinity of $\phi=0$, it should go to the state of true vacua $\phi=\pm v$. The barrier between them, however, forbids a classical rolling down to the true vacua, so the transition proceeds via quantum tunneling through the nucleation of bubbles of the true vacuum, whose profile is obtained by solving the field equation in the Euclidean space~\cite{Coleman:1977py}.
Since neither of the true vacua $\phi=\pm v$ has deeper free energy than the other, in principle the scalar $\phi$ would stay, in the equal probability, at the state of $\phi=v$ or $\phi=-v$ in the interior of the bubble. This results in the nucleation of the opposite-vacuum bubbles, denoted as $\phi_+$ and $\phi_-$ bubbles, from the first-order PTs.

\begin{figure}[t]
\includegraphics[width=0.48\textwidth]{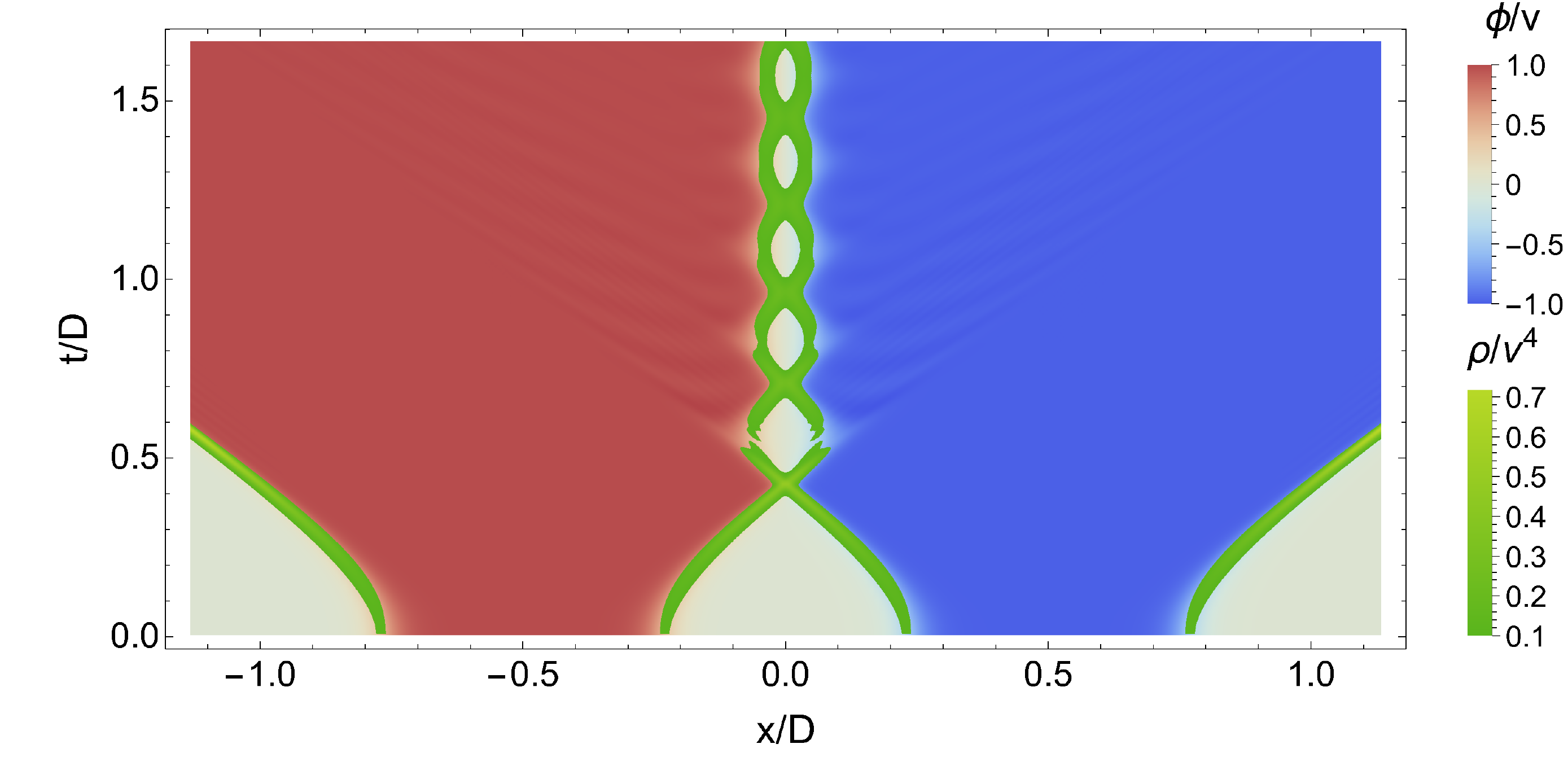}\\
\hspace*{0 mm}
\includegraphics[width=0.47\textwidth]{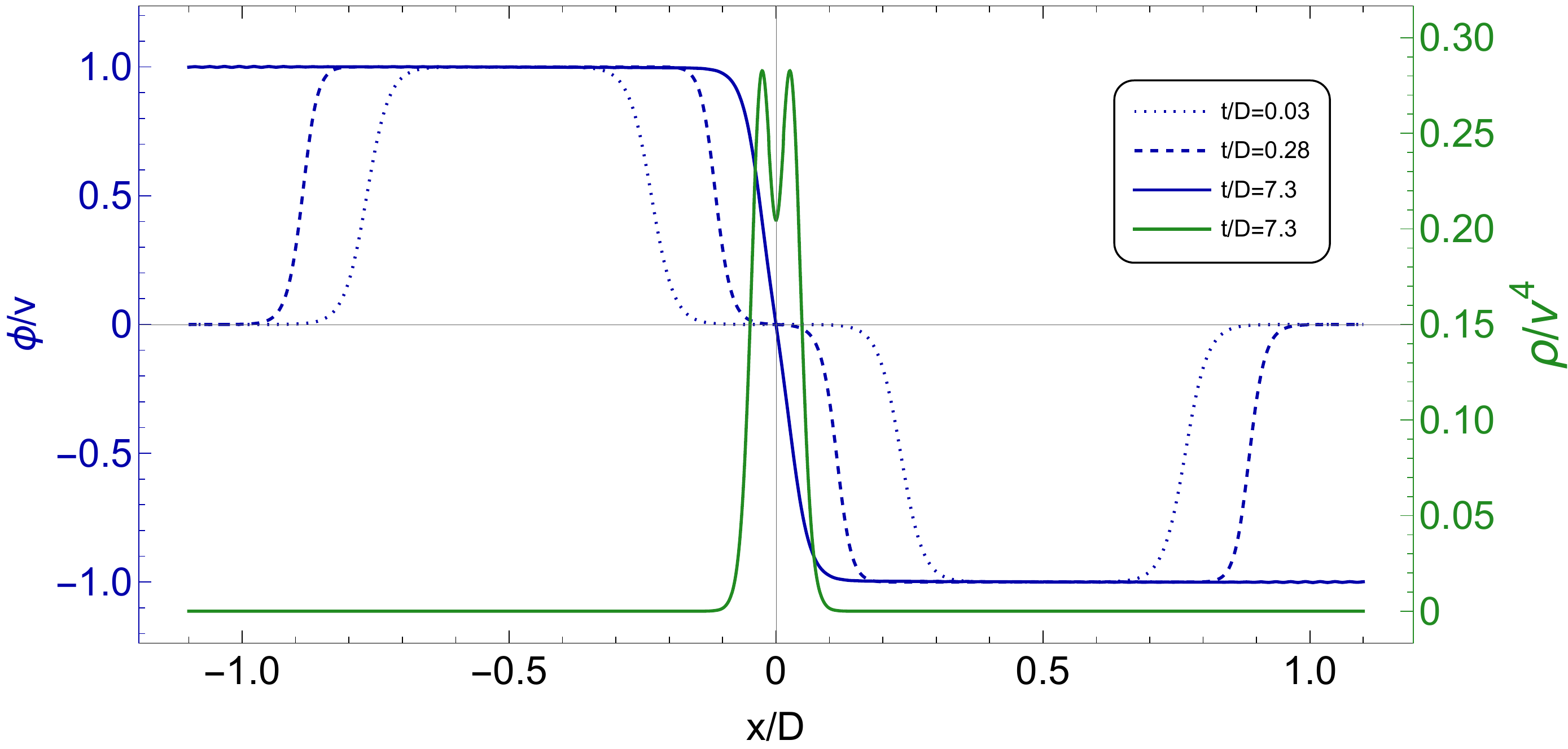}
\caption{Top: Evolution of two bubbles of the opposite vacuum that are initially separated by a distance $D$. $\phi_+$ and $\phi_-$ bubbles are respectively drawn in red and blue and their bubble centers point to the $x$ axis. The green color band stands for the energy density of the scalar field. Bottom: The field configuration along the $x$ axis at different times (blue lines) and the energy distribution of the domain wall (green line) when it reaches a stable field solution. 
}
\label{fig:twobubble}
\end{figure}

We start from the state in which $\phi_+$ and $\phi_-$ bubbles are placed with a separation $D$ much larger than their identical sizes. The evolution process is illustrated in Fig.~\ref{fig:twobubble} where the $x$ axis connects the two bubble centers. 
There we observe that once formed, two bubbles start the expansion and shortly approach the stationary speed close to the speed of light. In space between two bubble centers, 
a smooth interpolation of the field configuration occurs and the majority of the scalar energy is localized to the exterior surface of two bubbles. 
When two bubbles collide, their walls effectively bounce off each other and get slowing down due to the surface tension. This leads to the energy that will be continuously lost. Consequently, a narrowing domain wall with a special {\it double-layer structure} is developed and eventually reaches the stable field solution until a sufficient amount of energy has been consumed. This implies that the energy of the scalar field deposited in the bubble wall cannot be completely converted into the energy of the domain wall, part of the energy will be lost by particle radiation~\cite{Matsunami:2019fss}.  
It is noticeable that in the whole process the field value at the collision point $x=0$ is zero with small numerical uncertainty. In fact, this is the natural field configuration interpolating two degenerate vacua for a $\mathbb{Z}_2$-symmetric potential.\footnote{If two bubbles of the same vacuum collide, then the field at the collision point $x=0$ either bounces back to the local minimum $\phi=0$ or remains oscillating around the true vacuum~\cite{Braden:2014cra}.}
More importantly, this solution supports the stability of the produced domain wall. 
The width of the domain wall and how long it bounces before stabilizing depends on the initial separation $D$ and the parameter $a$ determining the ratio of the barrier height to the depth of the true vacuum relative to the potential of the false vacuum.

\emph{Dynamical evolution of domain wall network}.
To investigate how the domain wall network evolves in time, one should understand the dynamics of the scalar field that obeys the equation of motion
\begin{equation}
\ddot{\phi}-\nabla^{2} \phi+\frac{\partial V(\phi)}{\partial \phi}=0,
\label{eq:eom}
\end{equation}
with the potential $V(\phi)$ given in \eq{eq:potential}.
In this work, we consider the phase transitions that last much shorter than the Hubble time $H^{-1}_{*}$. 
Hence, the cosmic expansion is negligible in the evolution process. 
We also neglect the gravitational effect on the self-interaction of domain walls.
 
Due to the nonlinear nature of \eq{eq:eom}, we employ lattice simulations to solve it. 
We set up the initial state in the way that $N_b=64$ vacuum bubbles\footnote{The bubble profile is solved by the static equation of motion for the field $\phi$ in the Euclidean spacetime. For the thin-wall bubbles it has a simple analytic expression.} (consisting of  $\phi_+$ and $\phi_-$ bubbles) simultaneously nucleate in a three-dimensional box with the size being a small fraction of one Hubble volume. This determines the initial value of the scalar field $\phi$ on each lattice site.
For each set of bubble initialization, we run the simulations where the scalar field $\phi$ are evolved on a three-dimensional lattice with a total of $N^3 (N=500)$ points using a Crank-Nicholson leapfrog algorithm, which follows~\cite{Cutting:2020nla}. The lattice spacing $\Delta x=0.2/v$ is chosen such that the wall structure is distinguishable and the time interval is set $\Delta t= 0.2  \Delta x $. The periodic boundary conditions are imposed given the fact that the universe is isotropic and homogeneous within the simulation scale.

\begin{figure}[t]
\centering
\includegraphics[width=0.48\textwidth]{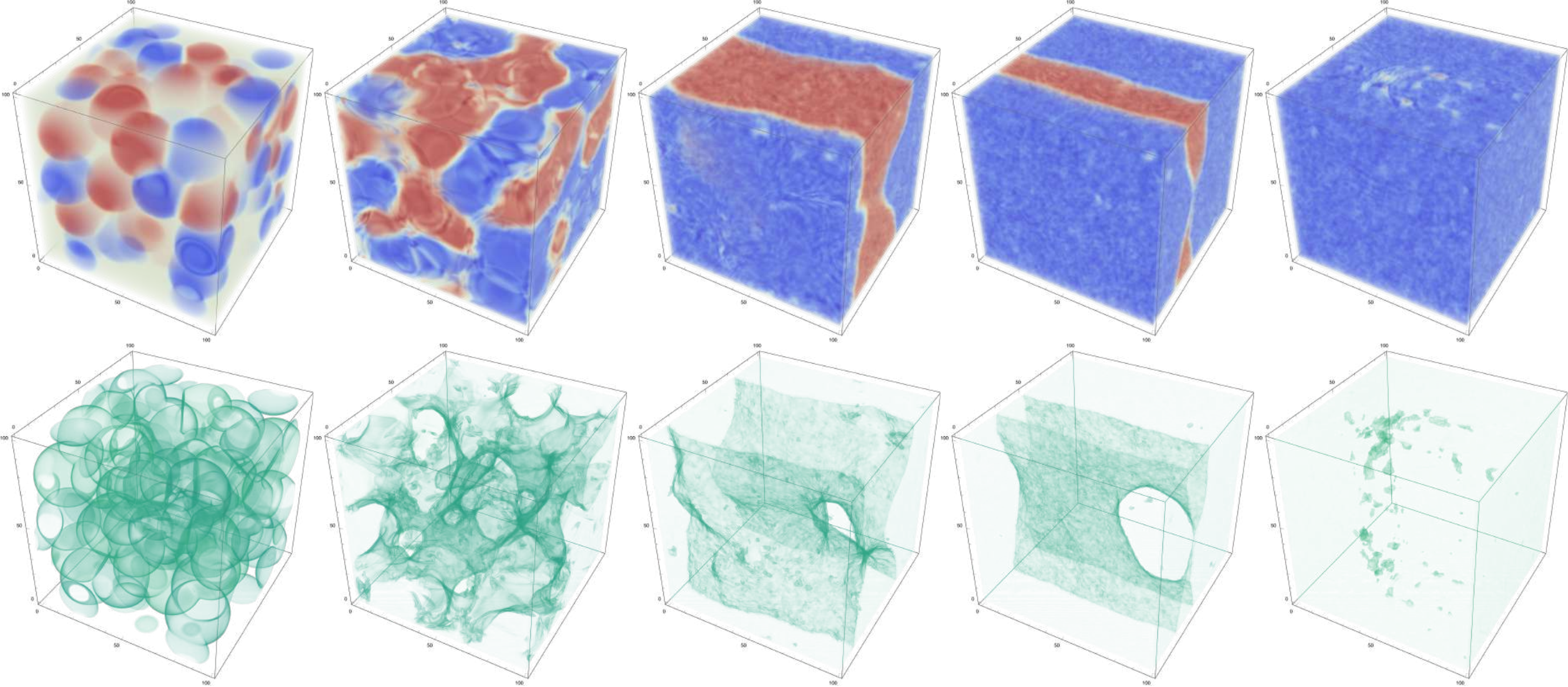}
\caption{Time evolution of the scalar field $\phi$ (upper) 
and the domain wall network (lower) produced from the first-order PTs. Color represents the value of the scalar field. The graphs from left to right respectively are taken at $t/R_*=0.5,1,3,5.6,7.8$. 
The first slice was taken when the bubbles collide, followed by a slice at $t/R_*=1$ where bubble percolate and the domain wall network is formed, which is close to the end of the phase transition. It ends up with 
a homogenous vacuum state in the simulation scale and the wall network completely decays away.
}
\label{fig:metadw}
\end{figure}

Fig.~\ref{fig:metadw} presents the time evolution of the scalar field $\phi$ and the domain wall network seeded by the vacuum bubbles of the first-order PTs. 
Shortly after the initial collision (first column) a series of small domain walls are formed in the region between $\phi_+$ and $\phi_-$ bubbles. As the bubbles continue expanding, they will absorb more regions that were resided in the old state of a metastable vacuum and coalescence proceeds among the bubbles of the same vacuum, causing the growth of long curved bubble walls. 
Consequently, a large network of the domain wall is developed  (second column). 
On the other hand, the wall network tends to collapse due to the tension of the wall curvature. As this effect dominates, the wall network completely decays away, perhaps very slowly.\footnote{The wall network is also possible to remain steady if it is stretched out in the simulation scale.} During this process the field configuration becomes more homogeneous on large scales, while the scalar field $\phi$ persists to oscillate on small wavelengths as long as the simulations run for times that are up to several $t/R_*$, eventually converting the simulation volume into either of the true vacua (i.e. $\phi_-$ vacuum in our simulation).

Nonetheless, Kibble mechanism~\cite{Kibble:1976sj} suggests that the configuration of the scale field $\phi$ is evolved in an independent way at separation exceeding a Hubble scale (cosmological horizon) and, in our example, can ends up with either $\phi_+$ or $\phi_-$ vacuum. Statistically, the probability of settling into these two types of vacuum state is the same, so it is unavoidable to form the domain walls on the cosmological scale, causing the domain wall problem still occurs. 

\begin{figure}[t]
\centering
\includegraphics[width=0.48\textwidth]{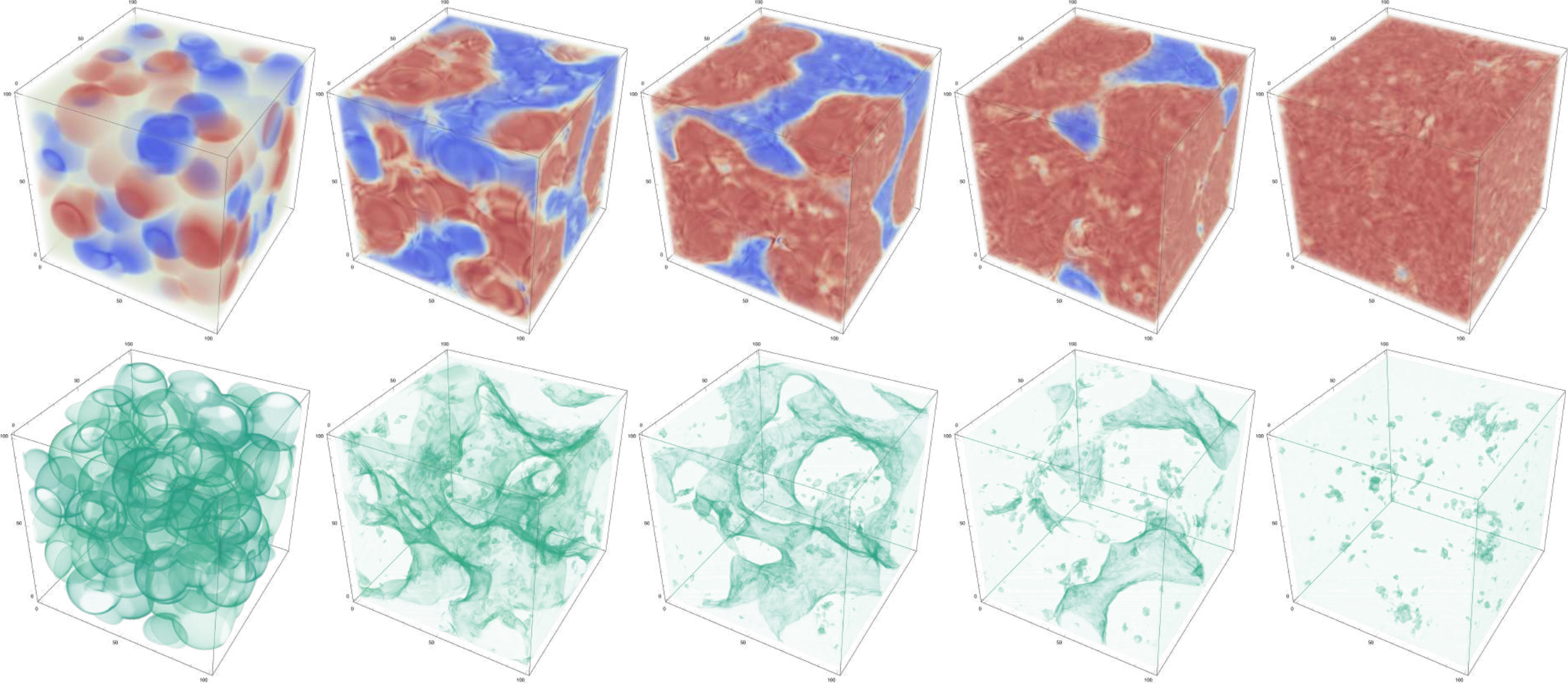}
\caption{Time evolution of the scalar field $\phi$ (upper) and the domain wall network (lower) with a bias potential $\epsilon = 0.01$ in the process until the domain walls completely collapse in the simulation volume. The slices from left to right are taken at $t/R_*=0.5,1,1.5,2,3.4$.}
\label{fig:collapsedw}
\end{figure}

To evade the problem we introduce a biased term in the potential~\cite{Saikawa:2017hiv}
\beq
\Delta V(\phi)=\epsilon v \phi\left(\frac{1}{3} |\phi|^{2}-v^{2}\right). 
\label{eq:biaspot}
\eeq
A non-zero value of $\epsilon$ breaks the degeneracy of the true vacua. But it cannot be too large otherwise the formation of domain walls will not happen at all. As an illustration, we take $\epsilon =0.01$, $\phi_+$ becomes the unique true vacuum while $\phi_-$ lifts to a metastable state.
From Fig.~\ref{fig:collapsedw} we observe that the impact of the bias term appears recognizable in the late evolution of the wall network.
While the wall network is still formed by the end of PT, it is no longer stretched to an infinite network but is quickly diluted and collapsed. Very soon the simulation region ends up with a homogenous state where the scalar field evolves to the unique true vacuum $\phi_+$ with small local fluctuations. 
These observations are consistent with our expectation that the bias term destabilizes the domain walls.
A larger pressure is exerted across the wall of $\phi_+$ bubbles, driving them to expand faster than the $\phi_-$ bubbles.
On the other hand, new $\phi_+$ bubbles are nucleated and submerged in the background of the $\phi_-$ vacuum state that has formed.\footnote{This possibility is not implemented in our simulation.}
Both effects will reduce the occupancy of the $\phi_-$ state, leading to the fragmentation of the wall network, which will collapse when the wall segment is too small to overcome the swallow by vast neighbor bubbles.

\begin{figure}[t]
\includegraphics[width=0.45\textwidth]{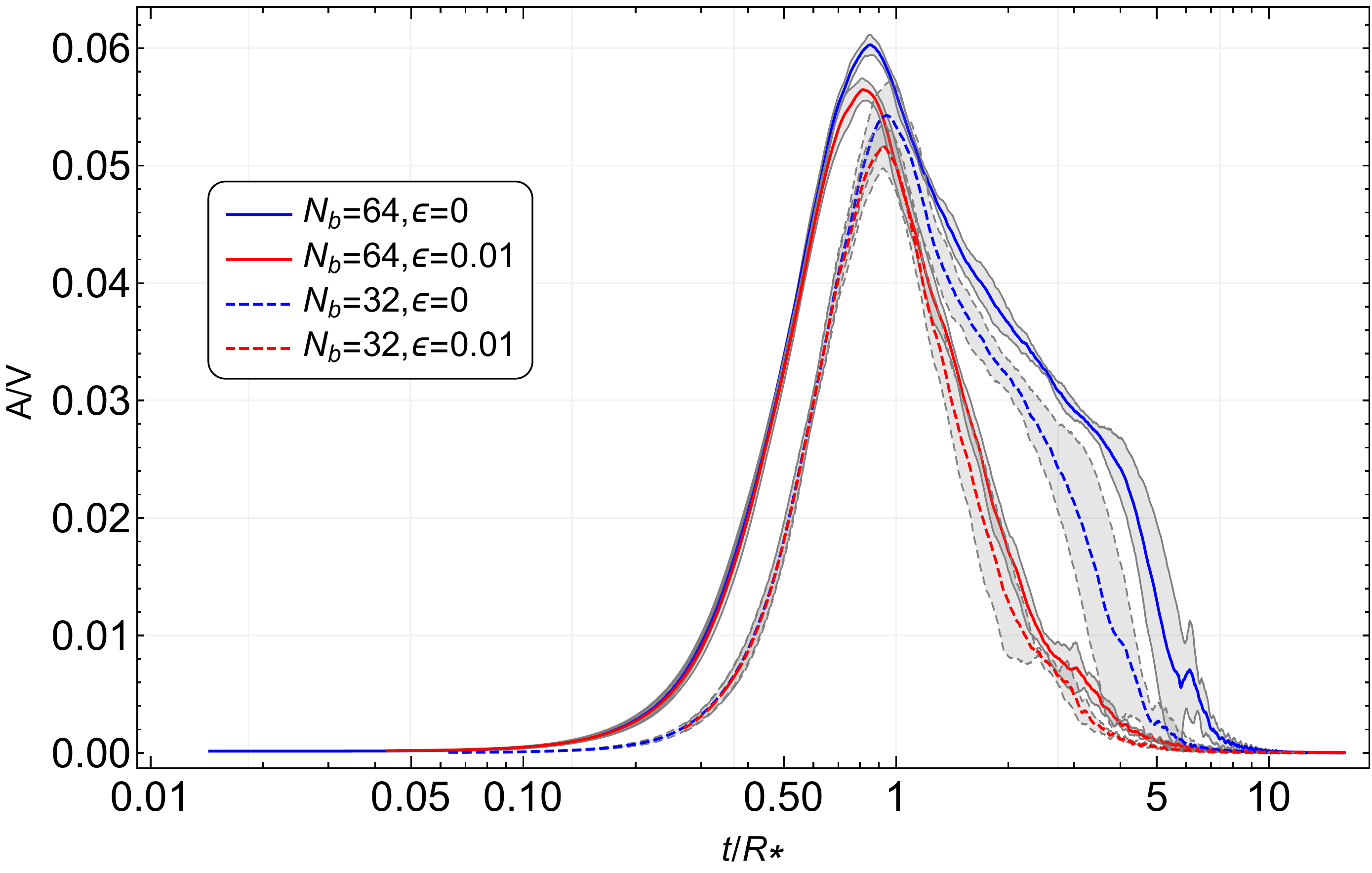}\\
\hspace{0.8mm}
\includegraphics[width=0.44\textwidth]{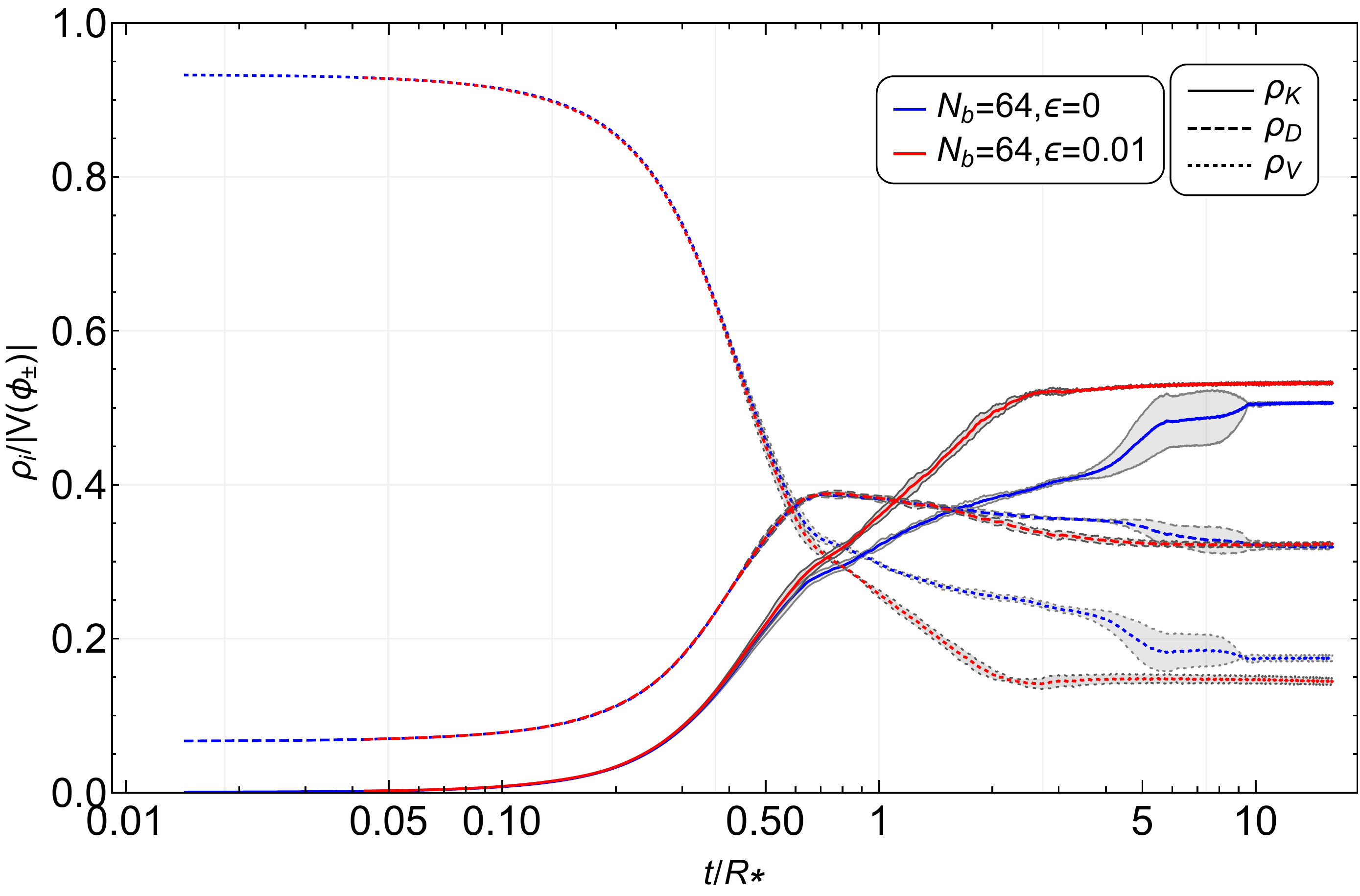}
\caption{Time evolution of the cross-sectional area of the domain wall network (upper) and the energy components of the scalar field (lower). Blue and red lines correspond to the wall network of the symmetric and nearly symmetric potentials. The error bars reflect how the scalar field evolves differently within the spatial regions in different simulations. In our simulations, the violation of the energy conservation is less than  $2.5\%$.}
\label{fig:dwarea_energy}
\end{figure}

Before moving to the main purpose of studying the GWs, we would like to quantitatively analyze how the bias term changes the dynamics of the wall network, through the time evolution of the cross-sectional area $A$ defined in~\cite{Press:1989yh}
and of the various components of the energy density of the scalar field 
\beq
\rho_K=\frac{1}{2}\dot{\phi}^2, \rho_D=\frac{1}{2} (\nabla \phi)^2, \rho_V=V(\phi)-V(\phi_\pm).
\eeq
The results of the normalized area $A/V$ and the normalized energy density $\rho_i/|V(\phi_\pm)|$ are presented in Fig.~\ref{fig:dwarea_energy}.  
We observe that 
the potential energy released from the PT $\rho_V$ converts to the kinetic energy $\rho_K$ and the gradient energy $\rho_D$ of the scalar field, making them increase as time goes, the latter one maximizes at $t/R_{*} \simeq 0.7$ which indicates the scalar fields distributed in the overlapping regions of the bubbles intensively oscillate following the initial collision. 
Meanwhile, the area in both cases maximizes around $t/R_{*}\simeq 0.8$ at which bubble collision is roughly complete, with a small suppression for the asymmetric potential ($\epsilon \ne 0$).

The main difference occurs in the decay process of the wall networks. With the bias term, although very small, the area of the wall network decreases exponentially in time. Typically, around $t/R_* \simeq 3$ the wall network decays away.
In contrast, the wall network of the symmetric potential is rather long-lived.
In both cases, each component of the energy density tends to be stable when the domain wall collapses due to large curvature tension, with a small difference in $\rho_V$. 
This occurs until $t/R_{*}\simeq 10$ for the symmetric potential and earlier at $t/R_{*}\simeq 5$ for the asymmetric potential.

In order to understand the variety of bubble configurations that affect the formation of the wall network, we also perform the simulations initialized with fewer bubbles of $N_b=32$. 
In this case, the mean separation distance between bubbles $R_*$ will increase.
This usually takes a longer time for bubbles to collide and percolate, therefore delaying the formation of wall network while accelerating the decay process. 
Notice that there is a rather large uncertainty for the symmetric potential, this is because of the fact that the domain wall network can remain steady or rapidly collapse after the formation.

\emph{The power spectrum of the produced gravitational waves}.
In the mechanism discussed above, the GW source is the anisotropic stress-tensor coming from both colliding bubbles and collapsing wall network, 
\beq
\label{eq:stress}
T_{i j} (t, {\bf x})=\partial_{i} \phi \partial_{j} \phi-g_{i j}\left[\frac{1}{2} g^{\rho \sigma} \partial_{\rho} \phi \partial_{\sigma} \phi-V(\phi)\right].
\eeq
This hybrid source generates stochastic background of GW
whose energy density is defined as~\cite{Cutting:2020nla}
\begin{equation}
\rho_{\mathrm{gw}}(t)=\frac{1}{64 \pi G}\left\langle\dot{h}_{i j}^{\text {TT}} \dot{h}_{i j}^{\text {TT}}+ \nabla h_{i j}^{\text {TT}} \nabla h_{i j}^{\text {TT}} \right\rangle_V, 
\end{equation}
where ${h}_{i j}^{\text {TT}} (t, {\bf x})$ is the transverse-traceless (TT) tensor perturbations to the flat background metric and $\langle \dots \rangle_V$ denotes the spatial average for random field $\phi$ in the entire simulation lattice. 
Normalized by the critical energy density of the universe $\rho_c=3H^2/ (8\pi G)$, we obtain the power spectrum of the dimensionless GW energy density parameter 
\begin{equation}
\Omega_{\mathrm{gw}}(t,k)=\frac{ k^{3}}{128 \pi^3 GV\rho_c}  \int \!d \Omega (\left|\dot{\tilde h}_{i j}^{\mathrm{TT}}(t, \mathbf{k})\right|^{2}+ k^2 \left| \tilde h_{i j}^{\mathrm{TT}}(t, \mathbf{k})\right|^{2} ).
\end{equation}
where $\tilde h^{\text {TT}}_{ij}(t, {\bf k})=\Lambda_{ij,lk} ({\bf k})\tilde u_{lk} (t, {\bf k})$\footnote{The TT projection operator $\Lambda_{ij,lk}  ({\bf k}) = P_{ik}  ({\bf k})  P_{jl}  ({\bf k})-1/2 P_{ij}  ({\bf k})  P_{lk}  ({\bf k})$, with $P_{ij}  ({\bf k})=\delta_{ij}- \hat{k}_i \hat{k}_j$.} is the TT component of an auxiliary tensor $\tilde u_{i j} (t, {\bf k})$ obeying the following equation~\cite{Garcia-Bellido:2007fiu,Figueroa:2011ye}
\beq
\label{eq:einsteineq}
\ddot{\tilde u}_{i j} (t, {\bf k}) - k^2 \tilde u_{i j} (t, {\bf k}) =16 \pi G \widetilde T_{i j} (t, {\bf k}).
\eeq

\begin{figure}[t]
\centering
\includegraphics[width=0.46\textwidth]{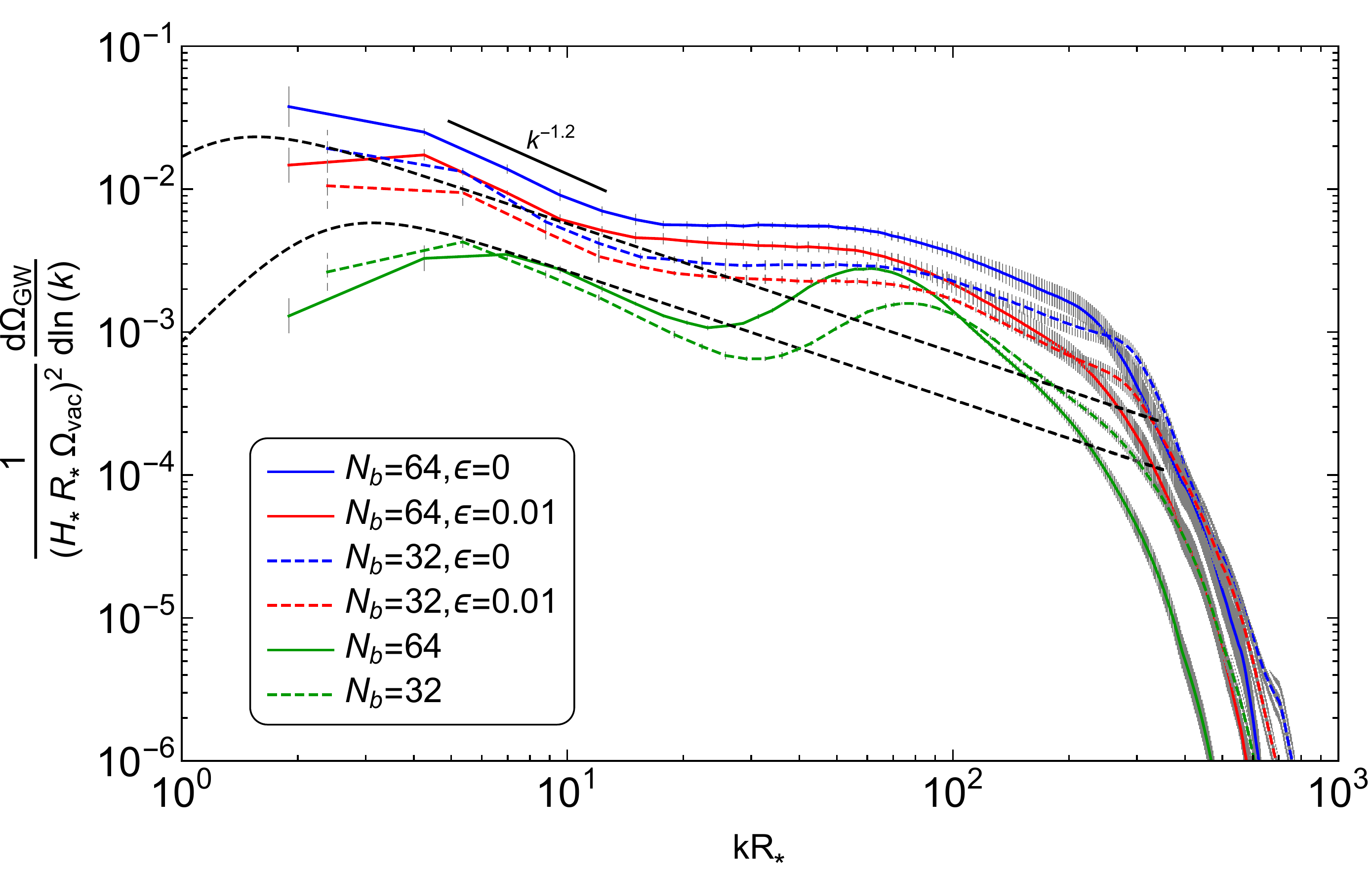} 
\caption{The gravitational wave power spectrum produced at $t/R_{*}=7.8$ in three models: bubble seeding domain wall (blue), bubble seeding metastable domain wall (red), and bubble-only (green) with the initial bubble numbers of $N_b=32$ (dashed lines) and $N_b=64$ (solid lines). The vertical bars indicate the spectrum variation obtained from the five simulations. 
The black solid line shows a decaying power law of $k^{-1.2}$. The power spectra predicted by the envelope approximation are shown in the black dashed lines for the parameter $\beta(R)=(8 \pi)^{1 / 3} v_{\text w}/R$ with $v_{\text w}=0.97$ and $R=R_*$ (lower), $2R_*$ (upper).}
\label{fig:gwspec}
\end{figure}

We summarize in Fig.~\ref{fig:gwspec} the resulting power spectra of GWs that are produced in three different models and with different initial bubble numbers. 
Each spectrum is generated by taking the ensemble average over the five field configurations with the different initial placement of bubbles. One can see that the wall networks, whether or not formed in the presence of the bias term, give broadly similar power spectra, while the wall networks from the symmetric potential leads to a higher amplitude spectrum due to the longer lifetime. 
In general, both power spectra are peaked near $k_{\text p} \simeq \mathcal{O}(1) \times \pi /R_*$ associated with the average bubble separation $R_*$, and have a knee around $k_{\text n}\simeq \mathcal{O}(10^2) /R_*$ with a more steep decay than $k^{-3}$ at the UV, which is difficult to explain as far as we know. Within the range, $kR_* \simeq \mathcal{O}(3-5) - \mathcal{O}(10^2)$ the power law is close to $k^{-1.2}$ followed by a slowly decreasing plateau. 
In addition, the IR power law appears a substantial departure from the causality law that requires a rise of $k^3$. However, due to the lack of the sufficient dynamic range in our simulations, this is just visible with relatively large error bars in the first two bins, making us difficult to draw a conclusive statement about the IR. 

\begin{figure}[t]
\centering
\includegraphics[width=0.45\textwidth]{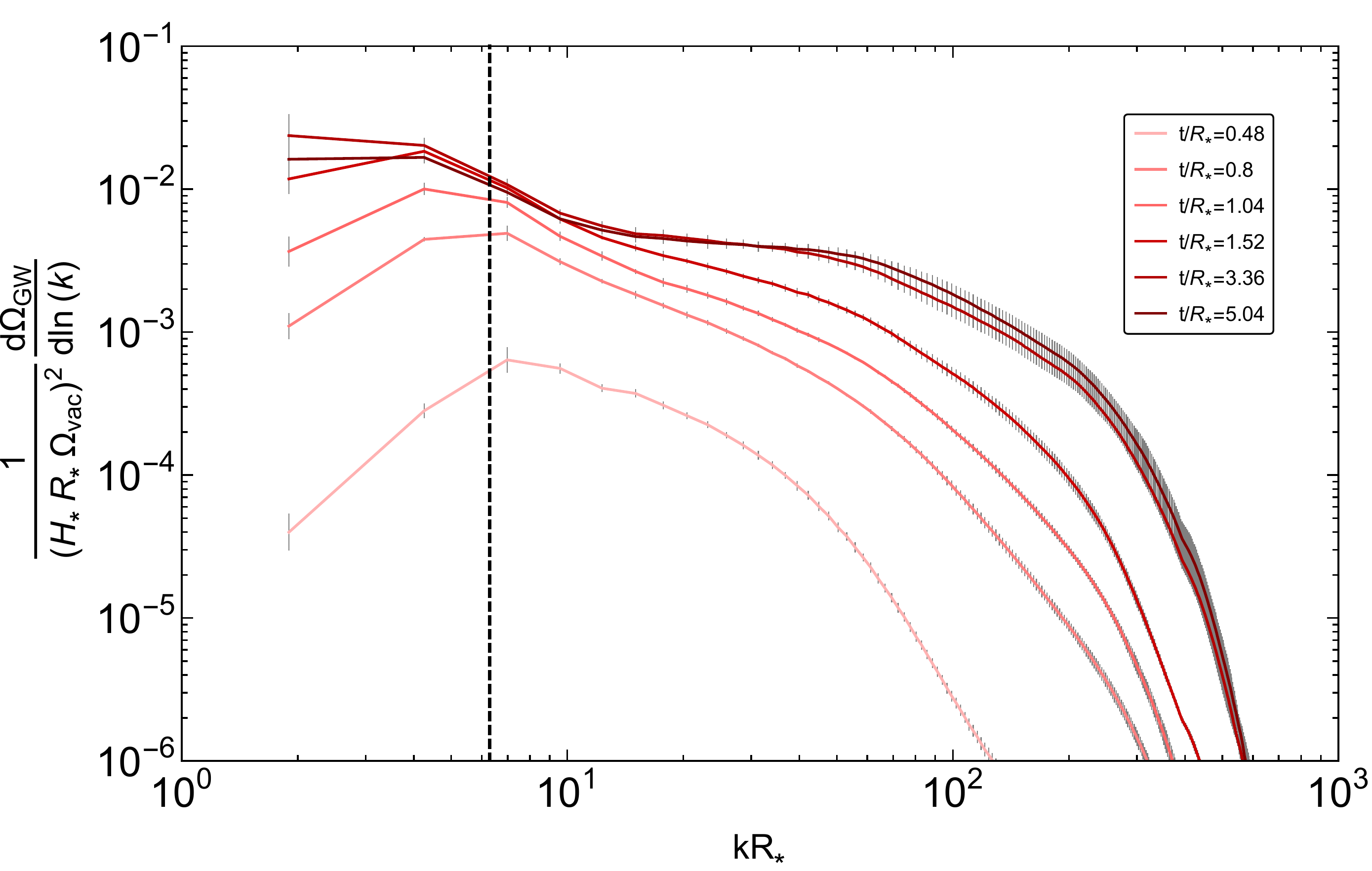} 
\caption{Time progress of the gravitational wave power spectrum from early bubble collision until the domain wall network completely decays away for the case of $\epsilon= 0.01$. We plot as a vertical dashed line the wave number $k = 2 \pi /R_*$.}
\label{fig:gwprogress}
\end{figure}

It is remarkable that the spectral shape of their power spectra is very different from the one sourced by the bubbles only. The latter is characterized by a double-peak structure, one associated with the average bubble separation $kR_*=2\pi$ close to the IR, and the other at high $k$ determined by the wall thickness $L_w$~\cite{Cutting:2018tjt}. However, for the models where the wall networks exist the UV bump that arises from the oscillation of the scalar fields in the overlapping wall regions~\cite{Cutting:2018tjt} is absolutely invisible in the whole process, as will be seen in Fig.~\ref{fig:gwprogress}.
This is perhaps because the UV bump associated with a much smaller length scale (perhaps the domain wall width) is buried by the strong growth of the UV spectrum that receives the substantial contribution from the wall networks.  
By comparing with the bubble-only case, we conclude that the GW power spectra receive the significant contribution from the wall networks, and this enhancement in the amplitude is roughly one order of magnitude at the primary peak.

In Fig.~\ref{fig:gwspec} we also include the result predicted by a semi-analytical model --- envelope approximation~\cite{Kosowsky:1992vn}.
Apparently, it cannot well describe the power spectra produced from two wall network models in the aspects of both spectral shape and amplitude. The discrepancy becomes much poorer as the frequency goes to the UV. Such inconsistency (or actually the underestimate) may be caused by too realistic modeling of the relevant GW sources. In fact, neither the bubble nonlinear oscillation during the PT nor the formation and decay process of the wall network after the PT is considered in the envelope approximation.
But we find that, for the model where the biased potential is included, the produced spectrum near the IR peak is consistent with the prediction by the envelope approximation, assuming the parameter that characterizes the PT duration $\beta(R)=(8 \pi)^{1 / 3} v_{\text w}/R$ is determined by $R=2R_*$. Such consistency may lead to the several crucial indications that the wall networks, before collapsing, kinematically proceed like the thin-wall bubbles, the metastable vacuum state $\phi_-$ is completely absorbed about $t/R_*=2$, and the largest average distance between neighboring walls is close to $R^{\text {max}}_{\text {dw}} \simeq 3R_*$, giving a prediction of the spectrum peak at $kR_*\simeq \pi$.

For completeness, we examine the dependence of the initial bubble number $N_b$ on the produced GW power spectrum. From the solid and dashed lines for the three models, we find that the value of $N_b$ has no significant change on the spectral shape, but a larger value of $N_b$ will leads to a spectrum with 
the overall increasing amplitude and a blue-shifted peak, both effects are because of the instability and the quicker collapse of the wall network that has been discussed. 
Our final comment is placed on the assumption of isotropy and homogeneity. It would be broken provided that the large-scale wall networks exist and are long-lived.
In such case, our prediction should be regarded as an estimate on the background spectrum on which small fluctuations or exotic structures are possible, which requires additional studies.

Fig.~\ref{fig:gwprogress} shows how GWs are accumulated in time for the model of bubble seeding metastable domain wall with $\epsilon=0.01$. This may help us to understand the underlying mechanism of GW production from the relevant GW sources. 
We see that at early times $t/R_* \simeq 0.5$ the bubbles begin to collide (see Fig.~\ref{fig:collapsedw}) and the domain wall area reaches the half maximum (see Fig.~\ref{fig:dwarea_energy}), the GW power spectrum grows with a peak near $kR_* \simeq 2\pi$, with a power law fall-off towards the UV.
Until $t/R_*\simeq 0.8$ the PT is, more or less, complete, the GW spectrum always has strong growth, and the expected $k^{-1}$ decay law as predicted by the envelope approximation~\cite{Kosowsky:1992vn}, indicating that colliding bubbles are the dominant GW source during the PT.

Nevertheless, unlike the bubble-only case~\cite{Cutting:2018tjt}, the production of GWs has not ceased after the PT completes ($t/R_*\simeq 1$). With the steady growth of the GW spectrum the peak persists but shifts toward lower values of $k$. This implies that GWs are being sourced on scales larger than $R_*$. In fact, after the PT the contribution to GWs is almost entirely from the formation and decay of the wall network (see Fig.~\ref{fig:dwarea_energy}).
This is supported by another observation that at $t/R_* \simeq 3$ the spectrum reaches the largest peak amplitude and decreases as steeply as $k^{-1.2}$ followed by a very slowly decreasing plateau within the range $kR_* \simeq 10-10^2$ before the UV fall-off.
At later times, the power spectrum has a slight decline in amplitude and falls at a higher UV cutoff.

\emph{Conclusions and Discussions}.
In this work, we develop a mechanism for producing the domain wall from the first-order PTs. If the breaking of a discrete symmetry happens with the PT, then the PT will proceed through nucleating bubbles of different true vacuum states that have degenerate potential energy. 
Taking the $\mathbb{Z}_2$ symmetry as an example, we show that a smooth transition in the field configuration between the bubbles of the opposite-vacuum state, when they are close to each other, will be made and the majority of the energy released from the PT is localized within the wall of the bubbles, forming the domain walls that are featured by a double-layer structure. By means of three-dimensional lattice simulations, we find that a large wall network can be formed in the universe that was filled with a considerable number of different bubbles. It is also long-lived in spatial regions exceeding the Hubble horizon and thus leads to cosmic catastrophe. To destabilize the domain walls, a bias potential breaking the degeneracy of the true vacuum is introduced so that the lifetime of the wall network is shortened, which will affect the produced GW power spectrum.


We have evaluated the GW power spectrum produced in the above-discussed mechanism. Our results show that both spectral shape and spectrum amplitude are very different from the ones from a first-order PT without the formation of domain walls. 
In the presence of domain walls seeded by the vacuum bubbles, the produced GW power spectra are characterized by two length scales: the largest average domain wall separation $R_{\text {dw}} \simeq (2-3) R_*$ and the average domain wall width $L_{\text {dw}}\ll R^{\text {max}}_{\text {dw}} $ after the bubble collision completes. 
Within the range of these two scales $k R_* \simeq \mathcal{O}(3-5) - \mathcal{O}(10^2)$, the power spectrum has a characteristic $k^{-1.2}$ power law following a very slowly decreasing plateau. For large $kR_*$ values, the power law index is a bit smaller than $-3$, and for small $kR_*$ values, it is less clear, which requires a larger-scale simulation to resolve. All these features are inconsistent with the prediction by the envelop approximation and therefore need further investigation.
The difference in the spectral shape actually occurs after the bubbles finish colliding, indicating colliding bubble is subdominant and even completely negligible in the later evolution of the wall network where the wall network becomes the main GW source.
Compared to the bubble-only case, our GW power spectrum is roughly increased by one order of magnitude at the primary peak. The short lifetime of the wall network leads to a slightly suppressed amplitude and a higher UV cutoff.

In this work, we do not discuss the relation between the feature of the produced GW spectrum and the properties of the vacuum bubbles, which require a delicate study. In addition, other approaches to destabilize the domain wall networks may lead to different GW power spectrum. If we assume the first-order PTs occur at the electroweak scale, then the domain wall will not be formed in the vacuum background but in the hot plasma which consists of relativistic particle gas of multiple species. This may essentially change the formation and evolution of the wall networks, resulting in different GW power spectrum as well. We leave both issues for further studies.

\section{acknowledgements}
\vspace*{-2mm}
This work is supported by the National Key Research and Development Program of China (Grant No. 2021YFC2203002) and in part by the GuangDong Major Project of Basic and Applied Basic Research (Grant No. 2019B030302001). Y. J. is also funded by the Guangzhou Basic and Applied Basic Research Foundation (No. 202102021092), the GuangDong Basic and Applied Basic Research Foundation (No. 2020A1515110150) and the Sun Yat-sen University Science Foundation.

\bibliography{bibtex} 
\end{document}